\documentclass[10pt,a4paper]{article}

\usepackage{amsmath,amssymb,amsthm}
\usepackage{url}
\usepackage{placeins}
\usepackage{subfigure}
\usepackage{multirow}
\usepackage{epsfig}
\usepackage{natbib}
\usepackage{rotating}

\newcommand\ben{\begin{equation}}
\newcommand\een{\end{equation}}
\newcommand\bea{\begin{eqnarray*}}
\newcommand\eea{\end{eqnarray*}}
\newcommand\bean{\begin{eqnarray}}
\newcommand\eean{\end{eqnarray}}

\begin{document}
\author{Andrew Green\footnote{Contact: andrew.green2@lloydsbanking.com} and Chris Kenyon\footnote{Contact: chris.kenyon@lloydsbanking.com}}
\title{MVA: Initial Margin Valuation Adjustment by Replication and Regression\footnote{\bf The views expressed are those of the authors only, no other representation should be attributed.}}
\date{First Submitted, February 24, 2014,\\
This Version, \today}

\maketitle

\begin{abstract}
Initial margin requirements are becoming an increasingly common feature of derivative markets. However, while the valuation of derivatives under collateralisation \citep{Piterbarg2010a,Piterbarg2012a}, under counterparty risk with unsecured funding costs (FVA) \citep{Burgard2011a,Burgard2011b,Burgard2013a} and in the presence of regulatory capital (KVA) \citep{Green2014b} are established through valuation adjustments, hitherto initial margin has not been considered. This paper further extends the semi-replication framework of \citet{Burgard2013a}, itself later extended by \citet{Green2014b}, to cover the cost of initial margin, leading to \emph{Margin Valuation Adjustment (MVA)}. Initial margin requirements are typically generated through the use of VAR or CVAR models. Given the form of MVA as an integral over the expected initial margin profile this would lead to excessive computational costs if a brute force calculation were to be used. Hence we also propose a computationally efficient approach to the calculation of MVA through the use of regression techniques, \emph{Longstaff-Schwartz Augmented Compression (LSAC)}. 
\end{abstract}

\section{Initial Margin and Funding Costs}
Initial margin requirements are becoming an increasingly common feature of derivative markets. Central counterparties (CCPs) require their members to post collateral through several mechanisms including initial margin, variation margin, volatility buffers or bid-offer costs and through clearing member contributions to the default fund. Under the Basel proposal for bilateral initial margin between financial counterparties \citep{BCBS-261}, all non-cleared derivatives between such entities will be subject to a requirement for initial margin by 2019. 

The valuation of derivatives under collateralisation is now well established and this has led to the wide acceptance of OIS discounting for trades between counterparties that are supported by a CSA agreement. \citet{Piterbarg2010a,Piterbarg2012a} developed the theory of pricing collateralised derivatives and this was later extended to include counterparty risk by \citet{Burgard2011a,Burgard2011b,Burgard2013a} leading to the introduction of \emph{Funding Valuation Adjustment (FVA)} for uncollateralised derivatives. The semi-replication framework introduced by \citet{Burgard2013a} was subsequently extended to include \emph{Capital Valuation Adjustment (KVA)} by \citet{Green2014b}.

Intuitively initial margin must be funded as in most cases rehypothecation of the initial margin is not allowed.\footnote{Under \citep{BCBS-261} limited rehypothecation of initial margin is allowed. However, the treatment of partial rehypothecation is beyond the scope of this article.} However, this effect must be demonstrated in a mathematically consistent way with all other valuation adjustments. To this end, in section \ref{sec:MVAreplication} we extend the semi-replication framework used by \citet{Burgard2013a} and \citet{Green2014b} to include the funding cost of initial margin. This leads directly to a further valuation adjustment term \emph{Margin Valuation Adjustment (MVA)}. 

The methodology used to calculate the size of initial margin for a given portfolio is often based on a historical Monte Carlo simulation approach to calculating Value-at-Risk (VAR) or variants of it such as Conditional Value-at-Risk (CVAR)\footnote{Also known as \emph{Expected Shortfall}}. For example, LCH.Clearnet SwapClear uses a proprietary model called PAIRS based on CVAR \cite{LCHCHP2013a}. Historical simulation parameters vary between CCPs with differences in the length of look-back period, confidence interval and close-out period \citep{Cameron2011a,Rennison2013a}.

As will become clear from the form of the MVA term, it will be necessary to calculate an expected initial margin profile as a function of time. This will require the initial margin to be estimated inside a Monte Carlo simulation. The use of a ``brute force'' approach to this calculation would require the calculation of multiple historical VAR scenarios on every path in a Monte Carlo simulation leading to high computational costs. Hence in section \ref{sec:MVAregression} we apply the established technique of Longstaff-Schwartz regression \citeyearpar{Longstaff2001a} to provide a computationally efficient approach to performing this calculation. Regression is used to provide a fast method of valuing a portfolio of derivatives and therefore acts a portfolio compression technique. To retain accuracy when considering the large shocks inherent in VAR, the state space used to generate the regression must be augmented beyond that generated by the Monte Carlo simulation. In this paper we propose a simple approach to the augmentation that can be applied to portfolios containing purely linear instruments and hence cover the majority of derivatives subject to clearing today, while for portfolios containing more complex instruments the ``early start'' Monte Carlo approach of \citet{Wang2009a} can be used. We have named the combined approach \emph{Longstaff-Schwartz Augmented Compression (LSAC)}.

To assess the relative size of MVA we calculate it for a portfolios of US dollar interest rate swaps in section \ref{sec:numres}. The MVA is then compared top the FVA that would be calculate on the same portfolio assuming it was unsecured. 

\section{MVA by Replication}\label{sec:MVAreplication}
To include the cost of initial margin alongside Credit, Funding and Capital Valuation Adjustments we extend the semi-replication model presented in \citet{Green2014b}, inself an extension of \citet{Burgard2013a}. This paper uses the notation of \citet{Green2014b}with additions (table \ref{table:not}). The sign convention is that the value of a cash amount is positive if received by the issuer. As with \citet{Green2014b} we seek to find the economic or shareholder value of the derivative portfolio, $\hat{V}$. 

\begin{table}
\centering
\small
\begin{tabular}{|p{3.5cm}|p{10cm}|}\hline
{\bf Parameter} & {\bf Description}\\\hline 
$\hat{V}(t, S)$ & The economic value of the derivative or derivative portfolio\\
$V$ & The risk-free value of the derivative or derivative portfolio\\ 
$U$ & The valuation adjustment\\
$X$ & Collateral\\
$I$ & Initial Margin posted to counterparty\\
$K$ & Capital Requirement\\
$\Pi$ & Replicating portfolio\\
$S$ & Underlying stock\\
$\mu_S$ & Stock drift\\
$\sigma_S$ & Stock volatility\\
$P_C$ & Counterparty Bond (zero recovery)\\
$P_1;\ P_2$ & Issuer bond with recovery $R_1$; recovery $R_2$, note $R_1\ne R_2$\\
$d\bar{\beta}_S;\ d\bar{\beta}_C;\ d\bar{\beta}_X;\ d\bar{\beta}_K;\ d\bar{\beta}_I$ & Growth in the cash account associated with stock; counterparty bond; collateral; capital; initial margin. All prior to rebalancing.\\
$r;\ r_C;\ r_i;\ r_X;\ r_F;\ r_I$ & Risk-free rate; Yield on counterparty bond; issuer bond; collateral;  issuer bond (one-bond case); initial margin\\
$M_B;\ M_C$ & Close-out value on issuer default; Counterparty default\\
$\alpha_C; \alpha_i$ & Holding of counterparty bonds; issuer bond\\
$\delta$ & The stock position\\
$\gamma_S$ & Stock dividend yield\\
$q_S;\ q_C$ & Stock repo rate; counterparty bond repo rate\\
$J_C;\ J_B$ & Default indicator for counterparty; issuer\\
$g_B;\ g_C$ & Value of the derivative portfolio after issuer default; counterparty default\\
$R_i;\ R_C$ & Recovery on issuer bond $i$; counterparty derivative portfolio\\ 
$\lambda_C;\ \lambda_B$ & Effective financing rate of counterparty bond $\lambda_C= r_C - r$; Spread of a zero-recovery zero-coupon issuer bond. For bonds with recovery the following relation holds $(1 - R_i)\lambda_B = r_i - r$ for $i\in\{1,2\}$\\
$s_F;\ s_X;\ s_I$ & Funding spread in one bond case $s_F = r_F - r$; spread on collateral; spread on initial margin\\
$\gamma_K (t)$ & The cost of capital (the assets comprising the capital may themselves have a dividend yield and this can be incorporated into $\gamma_K (t)$)\\
$\Delta\hat{V}_B;\ \Delta\hat{V}_C$ & Change in value of derivative on issuer default; on counterparty default\\
$\epsilon_h$ & Hedging error on default of issuer. Sometimes split into terms independent of and dependent on capital $\epsilon_h = \epsilon_{h_0} + \epsilon_{h_K}$\\
$P$ & $P = \alpha_1 P_1 + \alpha_2 P_2$ is the value of the own bond portfolio prior to default\\
$P_D$ & $P_D = \alpha_1 R_1 P_1 + \alpha_2 R_2 P_2$ is the value of the own bond portfolio after default\\
$\phi$ & Fraction of capital available for derivative funding\\\hline
\end{tabular}
\caption{\label{table:not}A summary of the notation, which is also common with Green et al. (2014).}
\end{table}

The following derivation follows \citet{Green2014b} closely. The dynamics of the underlying assets are given by
\begin{align}
dS = & \mu_s S dt + \sigma_s S dW\\
dP_C = & r_C P_C dt - P_C dJ_C\\
dP_i = & r_i P_i dt - (1 - R_i)P_i dJ_B\quad\text{for}\quad i\in\{1,2\}
\end{align}
On default of the issuer, $B$, and the counterparty, $C$, the value of the derivative takes the following values
\begin{align}
\hat{V} (t, S, 1, 0) = & g_B(M_B, X)\\
\hat{V} (t, S, 0, 1) = & g_C(M_C, X).
\end{align}
The two $g$ functions allow a degree of flexibility to be included in the model around the value of the derivative after default but with the usual close-out assumptions,
\begin{align}
g_B = &(V - X)^+ + R_B(V-X)^- + X\nonumber\\
g_C = &R_C(V-X)^+ + (V-X)^- + X,
\end{align}
where $x^+ = \max\{x, 0\}$ and $x^- = \min\{x, 0\}$.

We assume the funding condition:
\begin{equation}\label{eq:bondfunding}
\hat{V} - X + I + \alpha_1 P_1 + \alpha_2 P_2 -\phi K = 0,
\end{equation}
where the addition of $\phi K$ represents the potential use of capital to offset funding requirements. Comparing this with \citet{Green2014b} we see that the initial margin $I$ is funded through the issuance of bonds. There is only one term in the equation corresponding to posting initial margin to the counterparty and there is no corresponding term in initial margin posted to the issuer as we have assumed that this margin cannot be rehypothecated. Of course in the case of a CCP no such initial margin would be posted to the issuer in any case. The growth in the cash account positions (prior to rebalancing) are
\begin{align}
d\bar{\beta}_S = & \delta (\gamma_S - q_S) S dt\\
d\bar{\beta}_C = & -\alpha_C q_C P_C dt\\
d\bar{\beta}_X = & -r_X X dt\\
d\bar{\beta}_K = & -\gamma_K (t) K dt\\
d\bar{\beta}_I = & r_I I dt,
\end{align}
where an additional cash account is now included for any return received on the initial margin that has been posted to the counterparty. 

Using It\^{o}'s lemma the change in the value of the derivative portfolio is
\begin{equation}
d\hat{V} = \frac{\partial \hat{V}}{\partial t}dt + \frac{1}{2} \sigma^2 S^2 \frac{\partial^2\hat{V}}{\partial S^2}  dt +\frac{\partial \hat{V}}{\partial S} dS + \Delta \hat{V}_B dJ_B + \Delta \hat{V}_C dJ_C.
\end{equation}
Assuming the portfolio, $\Pi$, is self-financing, its change in value is
\begin{equation}
\begin{split}
d\Pi = & \delta dS + \delta (\gamma_S - q_S) S dt + \alpha_1 dP_1 + \alpha_2 dP_2 + \alpha_C dP_C \\
&- \alpha_C q_C P_C dt - r_X X dt - \gamma_K K dt + r_I I dt.
\end{split}
\end{equation}
Adding the derivative and replicating portfolio together we obtain
\begin{align}
d\hat{V} + d\Pi = & \Bigg[\frac{\partial \hat{V}}{\partial t} + \frac{1}{2} \sigma^2 S^2 \frac{\partial^2\hat{V}}{\partial S^2} + \delta (\gamma_S - q_S) S \nonumber\\
& + \alpha_1 r_1 P_1 + \alpha_2 r_2 P_2 + \alpha_C r_C P_C - \alpha_C q_C P_C - r_X X - \gamma_K K + r_I I\Bigg] dt\\\nonumber
& + \epsilon_h dJ_B + \left[\delta + \frac{\partial \hat{V}}{\partial S}\right] dS + \left[g_C - \hat{V} - \alpha_C P_C\right] dJ_C,
\end{align}
where
\begin{align}\label{eq:ehdef}
\epsilon_h = & \left[\Delta\hat{V}_B - (P - P_D) \right]\\\nonumber
= & g_B - X + P_D - \phi K
\end{align}
is the hedging error on issuer default. 

Assuming replication of the derivative by the hedging portfolio, except on issuer default gives,
\begin{equation}
d\hat{V} + d\Pi = 0,
\end{equation}
We make the usual assumptions to eliminate the remaining sources of risk,
\begin{align}
\delta = & - \frac{\partial \hat{V}}{\partial S}\\
\alpha_C P_C = & g_C - \hat{V},
\end{align}
and this leads to the PDE
\begin{align}\label{eq:VhatPDE}
0 = & \frac{\partial \hat{V}}{\partial t} + \frac{1}{2} \sigma^2 S^2 \frac{\partial^2\hat{V}}{\partial S^2} - (\gamma_S - q_S) S \frac{\partial \hat{V}}{\partial S} - (r + \lambda_B + \lambda_C) \hat{V}\nonumber\\
& + g_C\lambda_C + g_B \lambda_B - \epsilon_h \lambda_B - s_X X - \gamma_K K + r \phi K + s_I I\nonumber\\
& \hat{V}(T, S) = H(S).
\end{align}
where the bond funding equation \eqref{eq:bondfunding} has been used along with the yield of the issued bond, $r_i = r + (1- R_i) \lambda_B$ and the definition of $\epsilon_h$ in equation \eqref{eq:ehdef} to derive the result,
\begin{equation}
\alpha_1 r_1 P_1 + \alpha_2 r_2 P_2 = rX - rI -(r + \lambda_B)\hat{V} -\lambda_B (\epsilon_h - g_B) + r \phi K.
\end{equation}
Note that this paper assumes zero bond-CDS basis throughout. 

Writing the derivative portfolio value, $\hat{V}$, as the sum of the risk-free derivative value, $V$ and a valuation adjustment $U$ and recognising that $V$ satisfies the Black-Scholes PDE,
\begin{align}
\frac{\partial V}{\partial t} + \frac{1}{2} \sigma^2 S^2 \frac{\partial^2 V}{\partial S^2} - (\gamma_S - q_S) S \frac{\partial V}{\partial S} - rV = & 0\nonumber\\
V(T, S) = & 0,
\end{align}
gives a PDE for the valuation adjustment, $U$,
\begin{align}
 \frac{\partial U}{\partial t} & + \frac{1}{2} \sigma^2 S^2 \frac{\partial^2 U}{\partial S^2} - (\gamma_S - q_S) S \frac{\partial U}{\partial S} - (r + \lambda_B + \lambda_C) U = \nonumber\\
& V \lambda_C - g_C \lambda_C + V \lambda_B - g_B \lambda_B + \epsilon_h \lambda_B + s_X X - s_I I + \gamma_K K - r \phi K\nonumber\\
& U(T, S) = 0
\end{align}
Applying the Feynman-Kac theorem gives,
\begin{equation}
U = \text{CVA} + \text{DVA} + \text{FCA} + \text{COLVA} + \text{KVA},
\end{equation}
where
\begin{align}
\text{CVA} = & -\int_t^T \lambda_C(u) e^{-\int_t^u (r(s) + \lambda_B(s) + \lambda_C(s)) ds}\nonumber\\
& \times \mathbb{E}_t \left[V(u) - g_C(V(u), X(u))\right] du\label{eq:intCVA}\\
\label{eq:intDVA}\text{DVA} = & -\int_t^T \lambda_B(u) e^{-\int_t^u (r(s) + \lambda_B(s) + \lambda_C(s)) ds}\mathbb{E}_t \left[V(u) - g_B(V(u), X(u))\right] du\\
\label{eq:intFCA}\text{FCA} = & -\int_t^T \lambda_B(u) e^{-\int_t^u (r(s) + \lambda_B(s) + \lambda_C(s)) ds}\mathbb{E}_t \left[\epsilon_{h_0}(u)  \right]du\\
\text{COLVA} = & -\int_t^T s_X(u) e^{-\int_t^u (r(s) + \lambda_B(s) + \lambda_C(s)) ds} \mathbb{E}_t\left[X(u)\right] du\nonumber\\
= & +\int_t^T s_I(u) e^{-\int_t^u (r(s) + \lambda_B(s) + \lambda_C(s)) ds} \mathbb{E}_t\left[I(u)\right] du\nonumber\\
\label{eq:intKVA}\text{KVA} = & -\int_t^T e^{-\int_t^u (r(s) + \lambda_B(s) + \lambda_C(s)) ds} \nonumber\\
& \times \mathbb{E}_t \left[(\gamma_K (u) - r(u) \phi) K(u)+ \lambda_B \epsilon_{h_K}(u)\right] du.
\end{align}
The COLVA term now contains a adjustment for the initial margin. However this will vanish if the rate received on the post initial margin is equal to the risk free rate. In fact the FCA term contains the margin funding costs as we will now demonstrate. 

Consider the case of regular close-out with the funding strategy of semi-replication with no shortfall on default as described in \citet{Burgard2013a}. In this case there are two issued bonds, a zero recovery bond, $P_1$, which is used to fund the valuation adjustment and a bond with recovery $R_2 = R_B$ with a hedge ratio given by the bond funding equation \eqref{eq:bondfunding}. Hence we have,
\begin{align}
\alpha_1 P_1 = & - U\\
\alpha_2 P_2 = & -(V - \phi K - X + I).
\end{align}
The hedge error, $\epsilon_h$, is given by
\begin{align}
\epsilon_h = & g_B + I - X - \phi K + R_B \alpha_2 P_2\\\nonumber
= & (1-R_B) \left[(V - X)^+ - \phi K + I\right]
\end{align}
Hence we obtain the following for the valuation adjustment,
\begin{equation}
U = \text{CVA} + \text{DVA} + \text{FCA} + \text{COLVA} + \text{KVA} + \text{MVA},
\end{equation}
where
\begin{align}
\text{CVA} = & -(1-R_C)\int_t^T \lambda_C(u) e^{-\int_t^u(r(s) + \lambda_B(s) + \lambda_C(s)) ds} \mathbb{E}_t \left[(V(u))^+\right] du\label{eq:intCVAEx}\\
\label{eq:intDVAEx}\text{DVA} = & -(1-R_B)\int_t^T \lambda_B(u) e^{-\int_t^u (r(s) + \lambda_B(s) + \lambda_C(s)) ds}\mathbb{E}_t \left[ (V(u))^-\right] du\\
\label{eq:intFCAEx}\text{FCA} = & -(1-R_B)\int_t^T \lambda_B(u) e^{-\int_t^u (r(s) + \lambda_B(s) + \lambda_C(s)) ds}\mathbb{E}_t \left[(V(u))^+  \right]du\\
\text{COLVA} = & -\int_t^T s_X(u) e^{-\int_t^u (r(s) + \lambda_B(s) + \lambda_C(s)) ds} \mathbb{E}_t\left[X(u)\right] du\nonumber\\
\label{eq:intKVAEx}\text{KVA} = & -\int_t^T e^{-\int_t^u (r(s) + \lambda_B(s) + \lambda_C(s)) ds} \mathbb{E}_t \left[K(u)(\gamma_K (u) - r_B(u) \phi)\right] du\\
\label{eq:intMVAEx}\text{MVA} = & -\int_t^T ((1-R_B)\lambda_B(u) - s_I(u))e^{-\int_t^u (r(s) + \lambda_B(s) + \lambda_C(s)) ds}\mathbb{E}_t \left[I(u)  \right]du
\end{align}
As expected, the MVA takes the form of an integral over the expected initial margin profile. In this expression we have grouped the change to the COLVA term with MVA as both are determined by an integral over the initial margin profile.

\section{Calculating VAR \emph{inside} a Monte Carlo Simulation}
\subsection{VAR and the Risk-Neutral Measure}
VAR is most commonly calculated using a historical simulation approach and hence the VAR scenarios that are generated lie in the real world measure. From equation \eqref{eq:intMVAEx} it is clear that to proceed we need to apply these shocks inside a risk-neutral Monte Carlo simulation. In this paper we choose to assume that the VAR shocks are exogenously supplied and that they do not change during the lifetime of the portfolio. This is equivalent to assuming a fixed VAR window. With this assumption the VAR at each state inside the risk-neutral Monte Carlo is simply a function of the shocks applied to the state generated by the Monte Carlo. Relaxing this assumption to allow the shocks to change inside the risk-neutral Monte Carlo would require an extended debate on combining the physical and risk-neutral measures and this lies beyond the scope of this article. 

\subsection{Longstaff-Schwartz Augmented Compression}\label{sec:MVAregression}
To make VAR and CVAR calculations efficient the revaluation of the portfolio needs to be very fast. The Longstaff-Schwartz regression functions provide a means to do this as each is simply a polynomial in the explanatory variables $O_i(\omega, t_k)$. Hence we can approximate the value of the portfolio in each of the scenarios by using the regression functions with the explanatory variables calculated using the shocked rates,
\begin{equation}
V^{\text{I}}_q \approx \bar{F}(\alpha_m, O_i(\bar{y}^a_q, t_k), t_k). 
\end{equation} 
It is important to note that we apply the Longstaff-Schwartz approach to all derivative products from vanilla linear products to more complex exotic structures. We also use the resulting regressions as a compression technique, that is, we seek one set of functions, $\bar{F}$, as an approximation of the whole portfolio value. Hence not only does the Longstaff-Schwartz approach replace each trade valuation with a polynomial but it replaces the need to value each trade individually thus giving further significant performance benefits. Given that only a single polynomial is used to replace the entire portfolio valuation, \emph{Longstaff-Schwartz Augmented Compression} provides a portfolio valuation cost that is constant and independent of portfolio size. Of course the regression phase of the calculation will itself be a function of the number of cash flows in the portfolio so the computational cost does grow with increased portfolio size. However, our results in the next section show that this is not a significant effect and that the computational costs is independent of portfolio size in practice.

\begin{figure}[htb]
\begin{center}
	\includegraphics[trim=0 0 0 0,clip,width=0.7\textwidth]{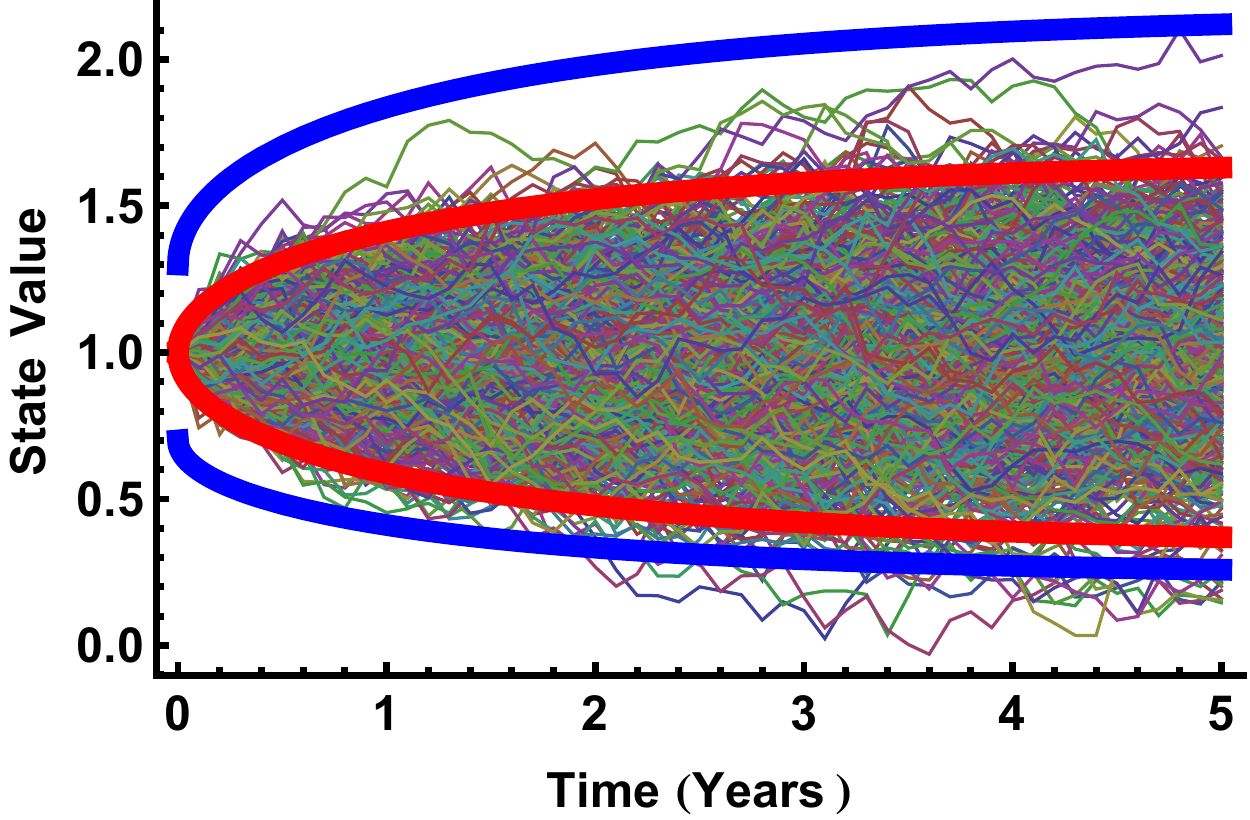}
\end{center}
\caption{\label{fig:scaleIMvar} Analysis of state space for an Orstein-Uhlenbeck process with parameters: $x0=1$; $\mu=1$; $\eta=1/4$; $\sigma=0.3$.  There are 1024 thin lines each representing one path.  Thick red lines show 1\%\ and 99\%\ confidence intervals for each date separately. Thick blue lines show those a 30\% relative shift applied to the confidence intervals. This magnitude of relative VAR shock was found in the 5-year time series used in the numerical examples in section \ref{sec:numres}.  To get a good regression approximation for the portfolio in region delimited by the thick blue lines the Monte Carlo alone is insufficient as few paths enter the region between the red and blue lines. Some form of augmentation is needed to generate a state space wide enough to give a good regression value for the portfolio once the VAR shocks have been applied.}
\end{figure}

Longstaff-Schwartz, in its original form, requires augmentation for the VAR and CVAR calculations because:
\begin{itemize}
	\item At $t=0$ portfolio NPV has exactly 1 value, so regression is impossible.
	\item For $t>0$ the state region explored by the state factor dynamics is much smaller than the region explored by VAR shocks.
\end{itemize}
To see why the state space generated by a Monte Carlo is not large enough when using regression for the portfolio value in the context of VAR, consider with a simple example where our model is driven by an Orstein-Uhlenbeck process (i.e. mean reverting), $dx = \eta(\mu -x)dt + \sigma dW,\ x(0)=x0$, where $W$ is the driving Weiner process.  Figure \ref{fig:scaleIMvar} shows the analysis of the state space with 1024 paths. It is clear that the 1024 paths shown do not cover the state space required by VAR calculation when the VAR shocks can give a shift of up to 30\% on a relative basis. This magnitude of relative VAR shock was found in the 5-year time series used in the numerical examples presented below in section \ref{sec:numres}. 

There are two augmentation methods that we can apply, \emph{early start Monte Carlo} and \emph{shocked state augmentation}.

\subsubsection{Early Start Monte Carlo}
This approach starts the Monte Carlo simulation earlier than today so that enough Monte Carlo paths are present in the region required to obtain accurate regression results for VAR shocks. \citet{Wang2009a} suggested the use of early start Monte Carlo in order to obtain sensitivities. The advantage of the early start Monte Carlo is that it preserves path-continuity. This is needed for portfolios which contain American or Bermudan style exercises. For such products a continuation value must be compared with an exercise value to obtain the correct valuation during the backward induction step in Longstaff-Schwartz. 

Given the portfolios we will consider in section \ref{sec:numres} contain only vanilla instruments we will not apply the early-start approach in this paper.

\subsubsection{Shocked-State Augmentation}

When portfolios do not contain American or Bermudan style exercise we can use a simpler method than Early Start to calculate regression functions giving portfolio values.   This is typically the case for central counterparties that do not deal with equity options. We call this approach Shocked-State Augmentation.  It is simpler because it does not have to preserve path-continuity of prices, as no backwards-induction step is required for valuation.  Thus the regression at each stopping date is independent of of all other regressions.  This also means they can be computed in parallel.

The objective of Shocked-State Augmentation is to have portfolio regressions that are accurate over the range of the state space relevant for calculation of VAR.  The state space relevant for VAR is strictly bigger than the state space explored by the simulation because VAR computation applies shocks to the state of the simulation, as illustrated in Figure \ref{fig:scaleIMvar}.  

Interpolation using a regression is much more likely to be accurate than extrapolation.  It is simple to construct regressions that are arbitrarily bad outside the range of their data.  Hence the idea in Shocked-State Augmentation is to expand the range of states at each stopping date, including t=0, so regressions interpolate rather than extrapolate.

The dimensionality of the state space {\it for VAR} at any stopping date on any path is given by the dimensionality of a VAR shock, not the driving factors of the simulation. We use ``VAR shock'' interchangeably with ``VAR scenario''. One VAR shock, for example, for a single interest rate may be described by 18 numbers giving relative movements of the zero yield curve at different tenors, and hence be 18-dimensional.  The fact that the simulation may be a 1-factor Affine interest rate model and hence describable by a 1-dimensional state is irrelevant.  The driving dimensionality is defined by the space explored by the VAR shocks.  Usually this will be larger than the dimensionality of the simulation model.  This high-dimensionality must be explored by augmenting the state space.  

At each stopping date the portfolio price is calculated as the sum of the component trades on each path.  If there are $m$ paths then this gives $m$ values to fit.  We can chose anything for the state variables (swaps and annuities in the example), and have as many as we like.  We fit a regression connecting the portfolio value to the stae variable values.  Usually $m<<n$ where $n$ is the dimensionality of a VAR shock so the problem is over determined.  We use a least-squares fit, so larger fitting errors are relatively highly penalized, under the assumption that these are more likely to occur with more extreme scenarios.  

We follow a parsimonious state augmentation strategy, that is complete at t=0.  We assume that there are more simulation paths than VAR shocks.
\begin{quote}
{\bf Shocked-State Augmentation:} apply one VAR shock at each stopping date, on each path.
\end{quote}
This strategy is parsimonious because we do not require any extra simulation paths, and because we use the same number of computations as for a usual simulation (apart from computing the effect of the shocks on the simulated data of course).  

This strategy is complete at t=0 in that we are certain to cover the full range required by VAR (as we have assumed more simulation paths than VAR shocks). This is automatic as we use the VAR shocks themselves to expand the state space.  Thus we are certain that all VAR computations will be within the range over which the regressions were calibrated.  So at t=0 we can expect any VAR computations to be close to exact provided we have sufficient basis functions.

Shocked-State Augmentation is the most parsimonious strategy in that it uses one VAR shock on each stopping date per simulation path.   In this version of Shocked-State Augmentation we pick the VAR shocks sequentially for each path at each stopping date,  So at t=1, say, path 1 uses shock 1, path 2 uses shock 2, etc.  As we have more paths than shocks we will use some shocks multiple times.

We are not interested in average effects of shocks --- VAR is an extreme result of the shocks on the portfolio.  Since the shocks cover a range of sizes (up to say 30\%\ relative) it is not obvious which direction in the n-dimensional space (defined by VAR shock dimensionality) will have the biggest effect on the portfolio.  For example, we cannot assumed that the appropriate direction is given by the local sensitivities (delta, gamma, vega, etc) of the portfolio.  Equally this is why we cannot simply pick the largest component of each of the VAR shocks and use this to expand the state space.  Although a shock defined as the maximum component of each shock would be large, we cannot say whether it is in the direction which changes the portfolio the most in n-dimensional space, for that magnitude of shock.  This is part of the need for the present technique.

In Shocked-State Augmentation the shocks are applied exactly as they would be for computing VAR.  In our experiments interest rates VAR shocks are multiplicative shocks on zero yield curve tenor points.  They are applied in Shocked-State Augmentation just as they would be for VAR to create a new market data state (at each particular stopping date on each path).

Many other strategies are possible, but  not covered for reasons of space.  We leave the optimal strategy for future research.  We compared our results to direct computation, i.e. no regressions and full revaluation (full results shown later).  For a reasonable number of basis functions the errors on outcome metrics are around 20bps of the notional or less.

\section{Numerical Results and Performance Comparison}\label{sec:numres}
\begin{figure}[ht]
\centering
\includegraphics[trim=0 250 0 275,clip,width=0.85\textwidth]{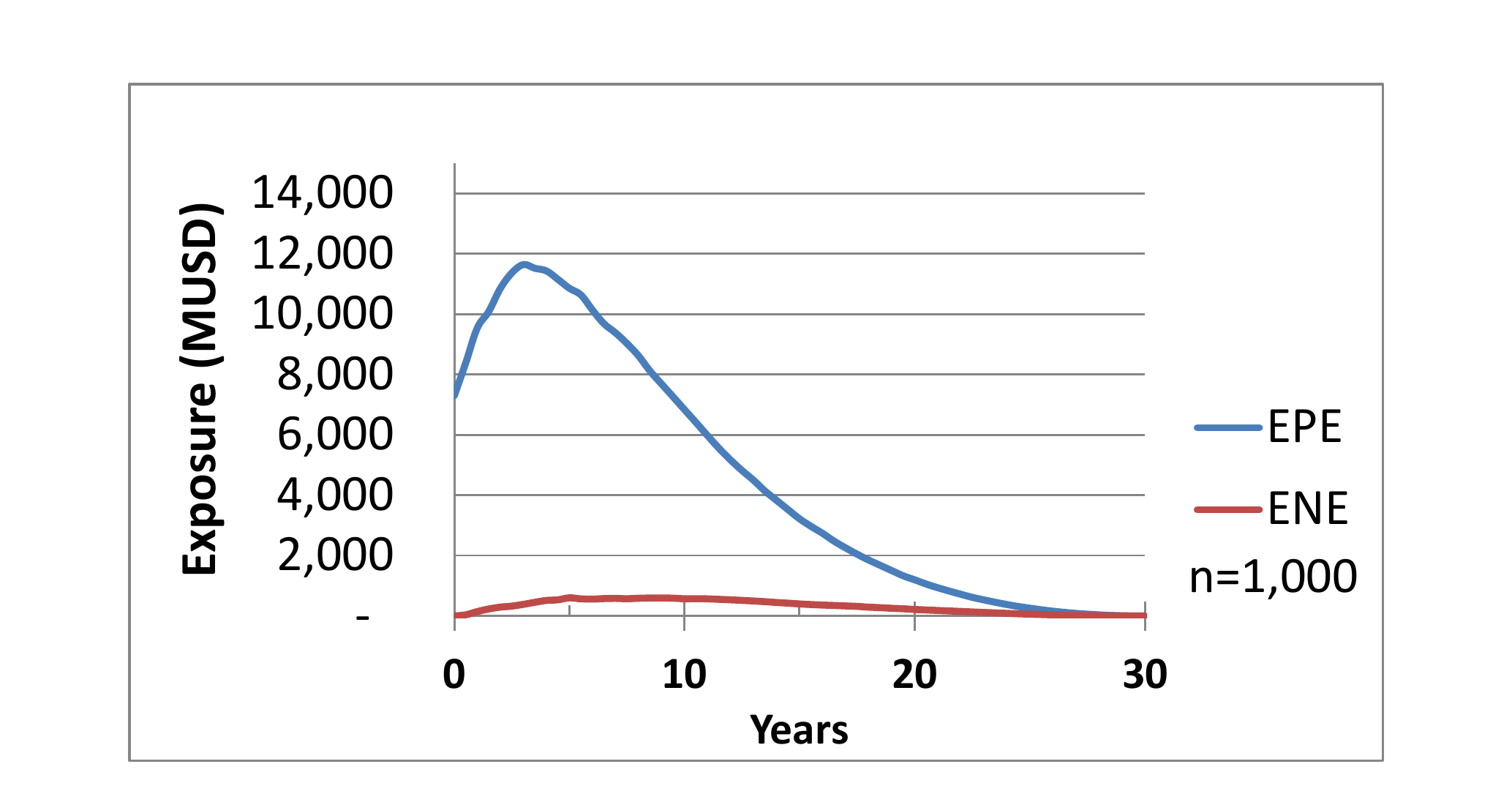}
\caption{\label{f:profile} The expected positive exposure (EPE, blue) and expected negative exposure (ENE, red) when $n = 1000$ and P[payer] = 90\%.}
\end{figure}

We calculate MVA on a series of portfolios of US Dollar interest rate swaps. We also calculate the FVA that would apply to the same portfolio if it were unsecured in order to provide a reference calculation to assess the impact of the MVA. We assume the use of the following initial margin methodology,
\begin{itemize}
	\item 99\%\ one-sided VAR;
	\item 10-day overlapping moves;
	\item 5-year window including a period of significant stress.  Our portfolio consists of IRS so a suitable period starts January 2007.  
	\item The 5-year window means that there were 1294 shocks.
\end{itemize}
Each VAR shock was a change to the zero yield curve.
\begin{itemize}
	\item Each VAR shock defined at 18 maturities: 0, 0.5 ,1 ,2 ,3 ,4 ,5 ,6 ,7 ,8 , 9, 10, 11, 12, 15, 20, 25, 30 years.
	\item Shocks are relative changes to zero yields, so given a zero yield $r$ at $T$ and a relative shock $s$  the resulting discount factor is:
	$	e^{-rT(1+s)}$.
	\item Linear interpolation in yield between shock maturities.
\end{itemize}
Test portfolios have $n$ swaps with maturities ranging up to 30 years, and each swap has the following properties,
\begin{itemize}
		\item $n$ swaps with maturity $i\times\frac{30}{n}$ where $i=1,\ldots,n$
		\item notional = USD100M $\times (0.5+x)$ where $x\sim$U(0,1) 
		\item strike = $K$ $\times (y+x)$ where $x\sim$U(0,1), $K=2.5\%$.  $y=1$ usually, or $y=1.455$ for the special case where we balance positive and negative exposures.
		\item gearing = $(0.5+x)$ where $x\sim$U(0,1)
		\item P[payer] = \{90\%, 50\%, 10\%\}
\end{itemize}
All swaps have standard market conventions for the USD market. For $n=1000$ the expected exposure profile of the portfolio is illustrated in Figure \ref{f:profile}.  We use $n=\{50,\ 100,\ 1000,\ 10000\}$ in our examples. The parameters MVA parameters used in the examples are as follows:
\begin{itemize}
	\item $\lambda_C = 0$ i.e. we assume that the issuer is facing a risk-free counterparty.
	\item $\lambda_B = 167bp$
	\item $R_B = 40\%$
	\item $s_I = 0$
\end{itemize}

The regression is performed using a linear combination of $2m+1$ basis functions including a constant, $m$ swaps, and $m$ annuities, with length $\frac{i}{m}\times 30,\ i=1,\ldots,m$. Note that we choose the basis functions once and use them for all stopping dates.  This choice was motivated by the fact that we can construct any swap {\it of the same maturity} but different fixed rates, from linear combination of a swap and an annuity and that we can construct forward-starting swaps from two swaps of different maturities.   We use 1024 paths in all our simulations with a horizon of 30 years and 6-monthly stopping dates.  The interest rate simulation model was calibrated to 10th March 2014.

\begin{figure}[htb]
\centering
\subfigure[As a function of basis size for a portfolio of $n=1000$ swaps.  Mean and max refer to errors in regressions across the range of stopping dates in the simulation between zero and 30Y.]{%
	\includegraphics[trim=0 250 0 275,clip,width=0.48\textwidth]{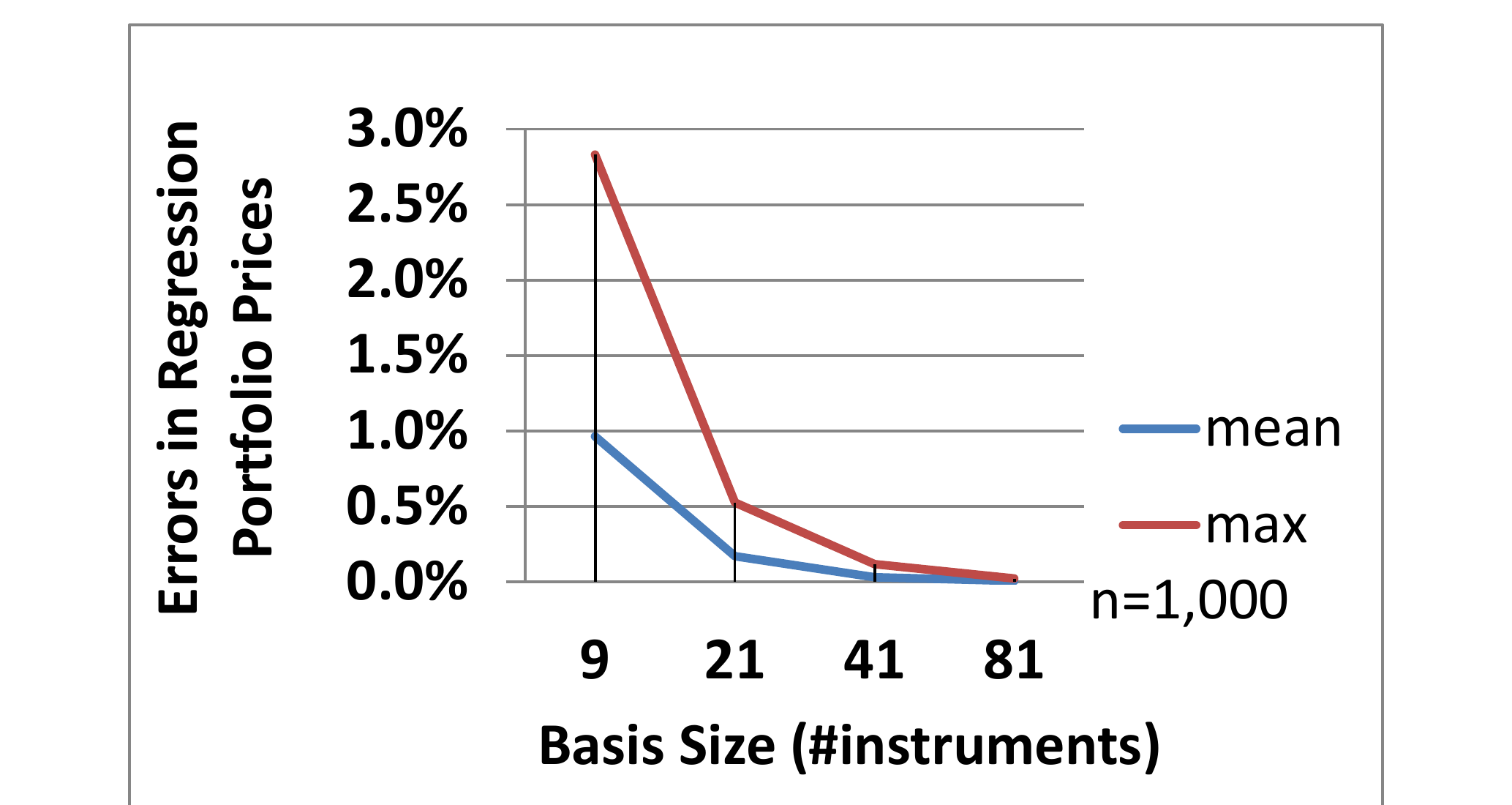}
  \label{fig:accuracy}}
\ 
\subfigure[Accuracy for portfolio metrics, as a function of portfolio size using 41 basis functions ($2\times20+1$), for the expected portfolio value, the 99\%\ one-sided VAR, and the IM funding cost.]{%
	\includegraphics[trim=0 250 0 275,clip,width=0.48\textwidth]{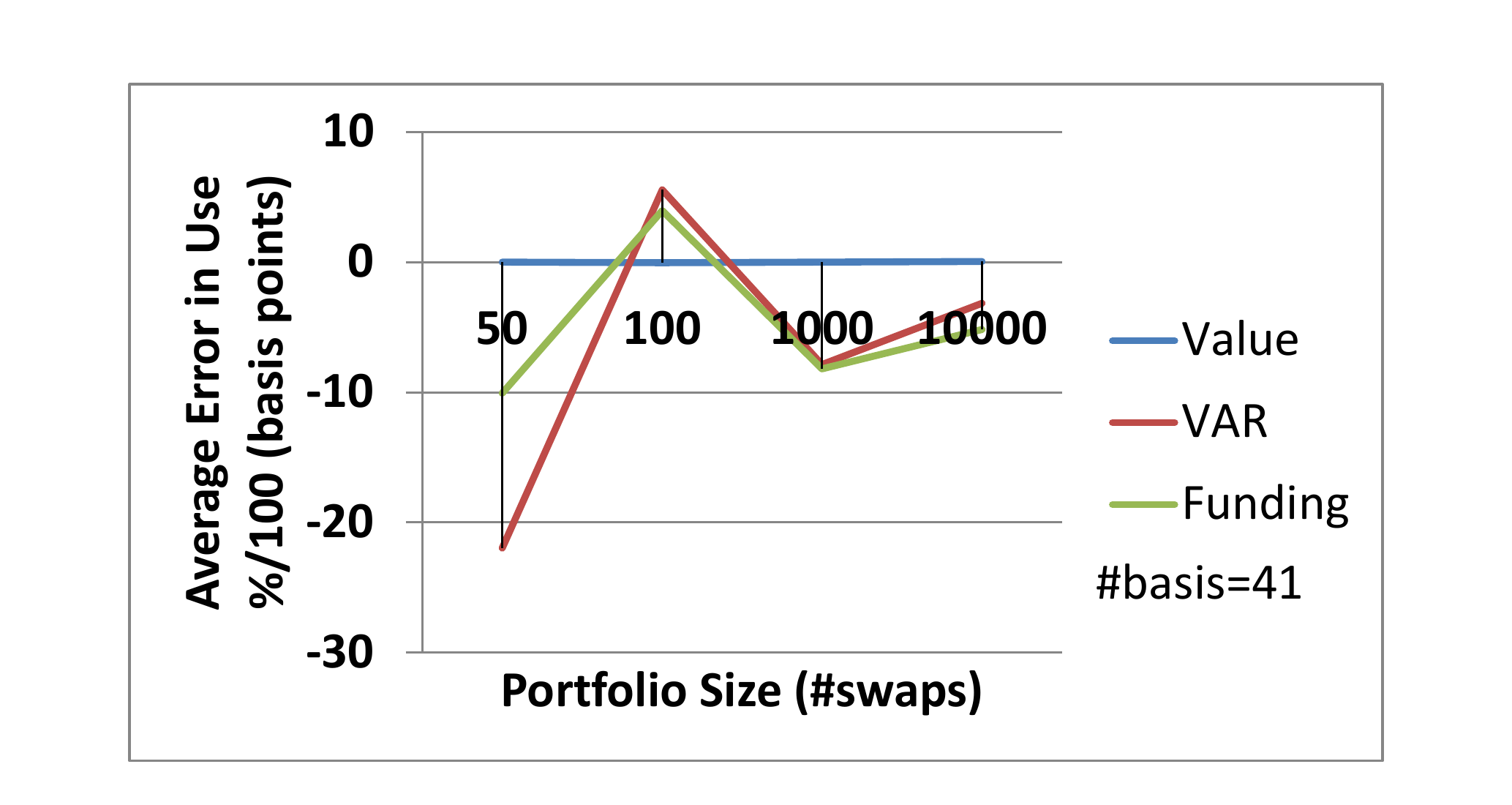}
	\label{fig:accuracyportfolio}}
\caption{Accuracy of LSAC method on test portfolios.}
\end{figure}

Regression accuracy on portfolio price is illustrated in Figure \ref{fig:accuracy}.   With a small number of basis functions high accuracy is achieved. Fewer basis functions are needed at later time points as the portfolio ages, this is handled automatically as we keep the same basis functions for all stopping dates.  As time progresses some of the basis functions mature, as does some of the portfolio.  The accuracy of the VAR, and IM funding, is shown in Figure \ref{fig:accuracyportfolio}.  

The algorithms were implemented in C++/CUDA and run on a GPU (NVIDIA K40c) for efficiency. This allowed us to perform brute force calculations for comparison purposes in a reasonable time frame. 

\begin{figure}[htb]
\begin{center}
	\includegraphics[trim=0 250 0 275,clip,width=0.95\textwidth]{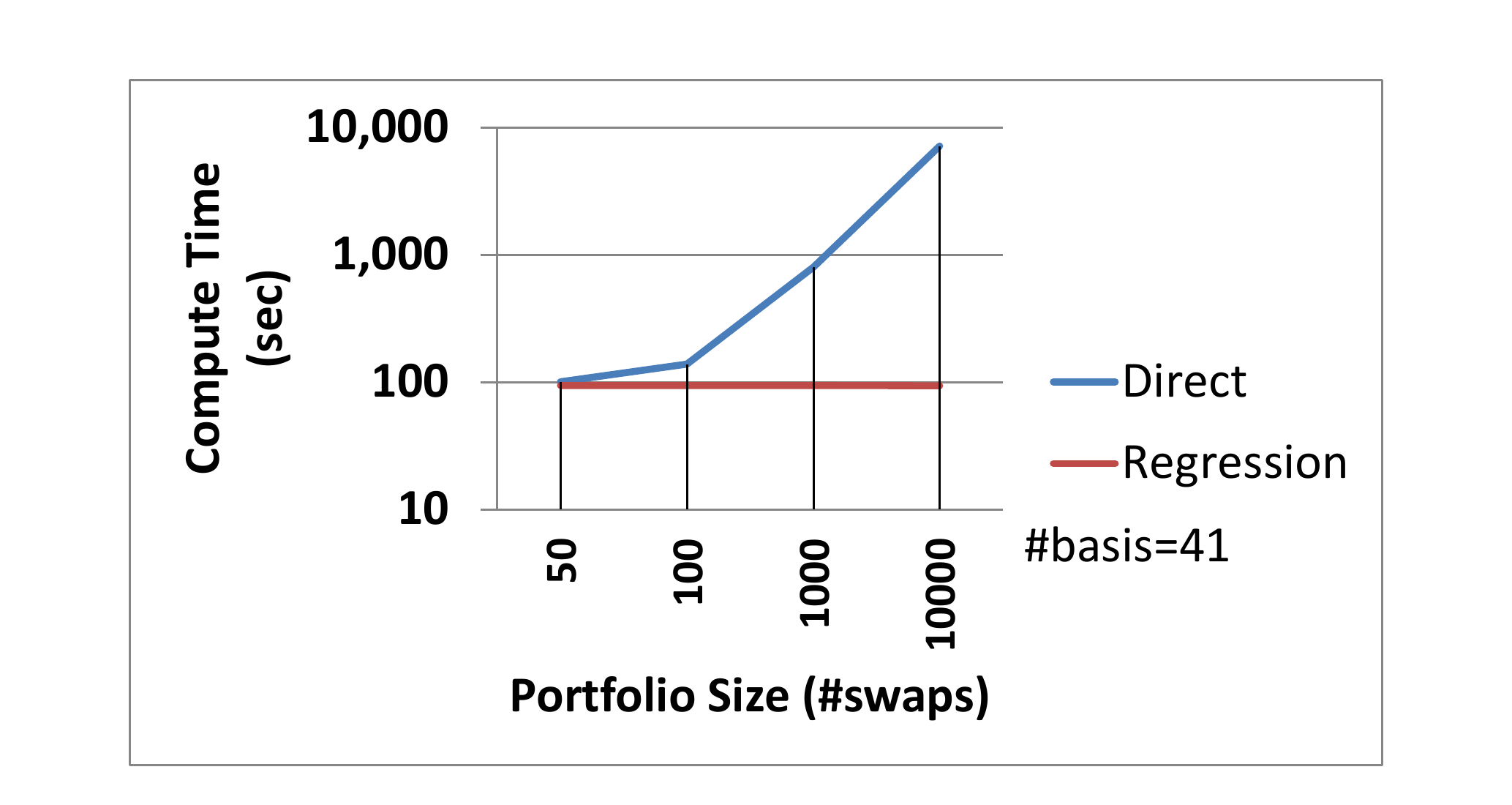}
\end{center}
\caption{\label{fig:portfoliotime} The performance of the LSAC approach (red) for the calculation of Initial Margin compared to the use of a brute force Monte Carlo within Monte Carlo (blue).  Note that both axes are logarithmic.}
\end{figure}

The performance of the LSAC approach is illustrated in Figure \ref{fig:portfoliotime} and is compared with a brute force calculation.  A brute force calculation means doing a full revaluation of the whole portfolio, i.e. repricing each trade, for each VAR shock at every stopping date on every path. Under the brute force approach the ES and VAR calculations take most of the time, particularly for larger portfolios. The regression approach, by contrast, is a constant-time algorithm for a specific number of VAR scenarios. The relative speedup increases with larger problem size, reaching x100 for medium-sized swap portfolio (10,000 swaps).

\begin{table}[htbp]
	\centering
		\begin{tabular}{ccc}
		Portfolio  & FVA & MVA \\
			P[payer]  & bps of notional & bps of notional \\ \hline
			90\%			& 115			& 53 \\
			50\%					& 0				& 2 \\
			10\%					& -113		& 56 
		\end{tabular}
	\caption{FVA and MVA costs relative to each other for the three example portfolios. Here $\text{FVA} = \text{FCA} + \text{DVA}$ is a symmetric approach to FVA costs and benefits.}
	\label{t:fva}
\end{table}
Table \ref{t:fva} shows the MVA cost for the three portfolios and compares this to the FVA. The MVA is close to 50\% of the cost of the FVA on the unsecured portfolio, and hence a significant valuation adjustment. 

\section{Conclusion}
This paper has extended the Burgard-Kjaer \citeyearpar{Burgard2013a} semi-replication approach to include the funding costs of initial margin and hence has added a further valuation adjustment, MVA. The form of MVA requires the expected initial margin profile to the calculated. Given that CCPs frequently use VAR or CVAR methodologies for initial margin, the calculation of expected initial margin would imply the need for a historical VAR simulation inside a risk-neutral Monte Carlo simulation. If a brute force approach were to be used this would be very computationally intensive. Hence we propose the use of Longstaff-Schwartz Augmented Compression to allow evaluation of the MVA using a computationally efficient algorithm that is largely independent of portfolio size. Computation of MVA for a number of example portfolios shows that is a significant adjustment and is just less than 50\% of the FVA of the same portfolio on an unsecured basis. 

\section*{Acknowledgments}
The authors would like to acknowledge the feedback from participants and reviewers at the GPU Technology Conference 2014, and Quant Europe 2014, where early results were presented. We also like to acknowledge the comments provided by the two anonymous referees. 

\bibliographystyle{plainnat}
\bibliography{kenyon_general}

\end{document}